\pdfoutput=1
\documentclass[aps, twocolumn, showpacs, superscriptaddress, prl]{revtex4-1}

\usepackage[pdftex]{graphicx}
\usepackage[pdftex]{epsfig}
\usepackage{epstopdf}
\usepackage{color}
\usepackage{amssymb,amsmath}
\usepackage[hypertexnames=false,colorlinks,urlcolor=blue,citecolor=blue]{hyperref}
\usepackage{braket}
\usepackage{siunitx}

\usepackage[dvipsnames]{xcolor}

\begin{document}

\title{Predictions of bound-bound transition signatures in x-ray Thomson scattering}

\author{A.D. Baczewski}
\affiliation{Center for Computing Research, Sandia National Laboratories, Albuquerque NM, USA}
\author{T. Hentschel}
\affiliation{School of Applied \& Engineering Physics, Cornell University, Ithaca NY, USA}
\author{A. Kononov}
\affiliation{Center for Computing Research, Sandia National Laboratories, Albuquerque NM, USA}
\author{S.B. Hansen}
\affiliation{Pulsed Power Sciences Center, Sandia National Laboratories, Albuquerque NM, USA}

\begin{abstract}
Bound-bound transitions can occur when localized atomic orbitals are thermally depleted, allowing excitations that would otherwise be forbidden at zero temperature.
We predict signatures of bound-bound transitions in x-ray Thomson scattering measurements of laboratory-accessible warm dense conditions.
Efficient average-atom models amended to include quasibound states achieve continuity of observables under ionization and agree with time-dependent density functional theory in their prediction of these scattering signatures, which hold compelling diagnostic potential for high-energy-density experiments.
\end{abstract}

\maketitle

\textit{Introduction.---}
The bulk properties of degenerate matter are determined by the Pauli exclusion principle~\cite{pauli1925zusammenhang}, in which no two indistinguishable fermions can simultaneously have the same quantum numbers.
For electrons bound to an atom or ion, which are typically described by quantum numbers for spin and orbital angular momenta, this has drastic spectroscopic implications.
For example, even if a photon that is resonant with a particular symmetry-allowed transition is available, it cannot be absorbed if the resulting electronic rearrangement would violate the exclusion principle.
However, this situation changes at non-zero temperatures in which there is a non-zero probability that any given electron is thermally excited.
Electronic rearrangements that would violate the exclusion principle at zero temperature become possible, giving rise to new spectral features.
These spectral features are commonly referred to as ``bound-bound transitions'' and they are well studied in the context of opacity measurements~\cite{bailey2015higher} where their sensitivity to temperature makes them effective thermometers.

Such thermometers are critical to the characterization of warm dense matter (WDM), the phenomenology of which is primarily determined by the presence of both degeneracy and thermal effects~\cite{graziani2014frontiers,dornheim2018uniform}.
Reliable estimates of temperature in WDM have long been confounded by the spatial and temporal non-uniformity of laboratory generated samples. 
This is compounded by the high electronic densities typical of WDM that make it opaque to optical probes.
Consequently, diagnostics that rely on scattering of hard x-rays are one of the most effective ways to characterize WDM, particularly with the increasing brilliance and sophistication of x-ray free electron lasers (XFELs)~\cite{sperling2015free,fletcher2015ultrabright}. 

Since the pioneering work of Glenzer \textit{et al.}~\cite{glenzer2003x}, x-ray Thomson scattering (XRTS) has been a workhorse diagnostic for WDM that includes electron and ion temperature among its sensitivities~\cite{doppner2009temperature}.
In an XRTS experiment, a beam of high-energy photons is directed at a sample of WDM and its energy-resolved scattering signature is measured at one or more angles.
Photons that scatter elastically from tightly bound electrons occupying the ions' core states generate momentum-dependent diffraction patterns that reveal ionic structure.
Inelastic scattering processes in which the photons exchange energy with collective or non-collective electronic degrees of freedom reveal details of the electron momentum distribution.
The breadth of processes probed by XRTS leads to numerous sensitivities including electron density~\cite{glenzer2007observations,lee2009x}, average ionization state~\cite{fletcher2014observations}, ion-ion correlations~\cite{ma2013x}, dynamic collision frequencies~\cite{faussurier2016electron}, conductivity~\cite{sperling2015free,witte2017warm}, and miscibility~\cite{frydrych2020demonstration}.

Interpretation of XRTS data relies on modeling the dynamic structure factor (DSF), S(q,$\omega$), which is proportional to the differential x-ray scattering cross section and parameterized by the momentum (q) and energy ($\omega$) transfer involved in the scattering process.
By finding the model inputs (i.e., physical variables) that predict the DSF that best agrees with the measured scattering cross section, one can infer the values of the physical variables that produced the data.
Broadly, models for the DSF can be separated into first-principles models that rely on approximating the inter-particle interaction (e.g., electronic exchange and correlation~\cite{baczewski2016x,mo2018first,ramakrishna2021first} or the absence of an attractive ionic potential~\cite{dornheim2018ab,moldabekov2021thermal});
single-center average-atom (AA) models~\cite{johnson2012thomson,souza2014predictions};
and simplified models~\cite{glenzer2003x} that rely on approximations to less fundamental quantities (e.g. the average ion charge $\langle Z \rangle$  ~\cite{murillo2013partial}).
While first-principles models have been used to assess sensitivities of the DSF to, e.g., temperature~\cite{mo2018first,ramakrishna2021first}, the computational cost of these models is too great for rigorous inference and the intrinsic systematic errors aren't yet sufficiently well-characterized for them to serve as a ground truth. 
Nevertheless, these models provide a critical mooring post in a sea of uncertainties and can validate other approximations made by simpler, less expensive models~\cite{baczewski2016x}.

In this Letter, we test the hypothesis that bound-bound transitions can contribute to the scattering signature that would be measured in an XRTS experiment.
To do so we make use of first-principles real-time time-dependent density functional theory (TDDFT), in which the primary approximation is the choice of a particular exchange-correlation potential~\cite{runge1984density-functional,marques2004time-dependent}.
We discover that bound-bound transitions are evident in systems like Al and Fe at experimentally accessible $\mathcal{O}(\SI{10}{\electronvolt})$ temperatures and can even come to dominate the scattering spectra as they do for opacities.
To facilitate the efficient analysis of these features, we extend an average-atom model to incorporate a bound-bound contribution to the Chihara decomposition~~\cite{chihara1987difference,chihara2000interaction} and develop a formalism for accommodating states near the continuum.
This new framework achieves good agreement with TDDFT and allows us to make projections for the diagnostic utility of the bound-bound contributions as novel thermometers for WDM.
We note that the bound-bound signatures discussed in this work are distinct from those recently observed in Resonant Inelastic X-ray Scattering (RIXS)~\cite{humphries2020probing}.
In terms of the Kramers-Heisenberg treatment of scattering~\cite{kramers1925streuung}, our predicted contributions arise due to the $|{\bf A}|^2$ contribution from the radiation field, rather than the ${\bf p}\cdot{\bf A}$ term in RIXS~\cite{kotani2001resonant}.
Thus these bound-bound transition signatures do not require resonance between the incident photon and matter under interrogation to be observed.

\textit{Methods---}
We use real-time TDDFT to compute the DSF without relying on the Chihara decomposition.
Since we are investigating bound-bound features that are absent in the traditional Chihara decomposition, their presence in a Chihara-free theory is a crucial confirmation that the effect we are seeking is represented.
We use the same computational methodology and its implementation as an extension to the Vienna \emph{ab initio} simulation package (\textsc{VASP}) \cite{kresse1996efficiency,kresse1996efficient,kresse1999ultrasoft} first described in Ref.~\cite{baczewski2016x}.
Starting from a Mermin-Kohn-Sham equilibrium state, the time-dependent Kohn-Sham equations are propagated in time in the presence of a sinusoidal scalar perturbing potential with a wave vector commensurate with the momentum transfer at which the DSF is to be computed.
This requires a supercell for which the desired wave vector is in the set of reciprocal lattice vectors.
The intensity of the perturbing potential is chosen to be sufficiently small that we remain in the linear response regime, and our results should be equivalent to those of energy domain formulations of the same problem~\cite{sakko2010time,ramakrishna2021first}.
However, we make use of the real-time formalism to maintain excellent strong scaling in our implementation~\cite{baczewski2016x}.
This is particularly important for the extreme conditions under consideration, which require the evolution of many partially occupied Mermin-Kohn-Sham orbitals~\cite{mermin1965thermal}.

We will see that TDDFT indicates the presence of bound-bound transitions in the DSF and have modified an AA model to include the relevant physics.
The model used here closely follows the methods developed by Starrett and Saumon~\cite{starrett2014hedp} and its application to XRTS~\cite{souza2014predictions}, but with three significant additions: a bound-bound term in the Chihara decomposition, a new treatment of quasibound states, and a Mermin dielectric function with an improved treatment of dynamical electron-ion collisions.
The latter two, in particular, resolve challenges in AA modeling that are somewhat independent of the topic of bound-bound transitions.

\begin{figure}[h]
    \centering
    \includegraphics[scale=0.8]{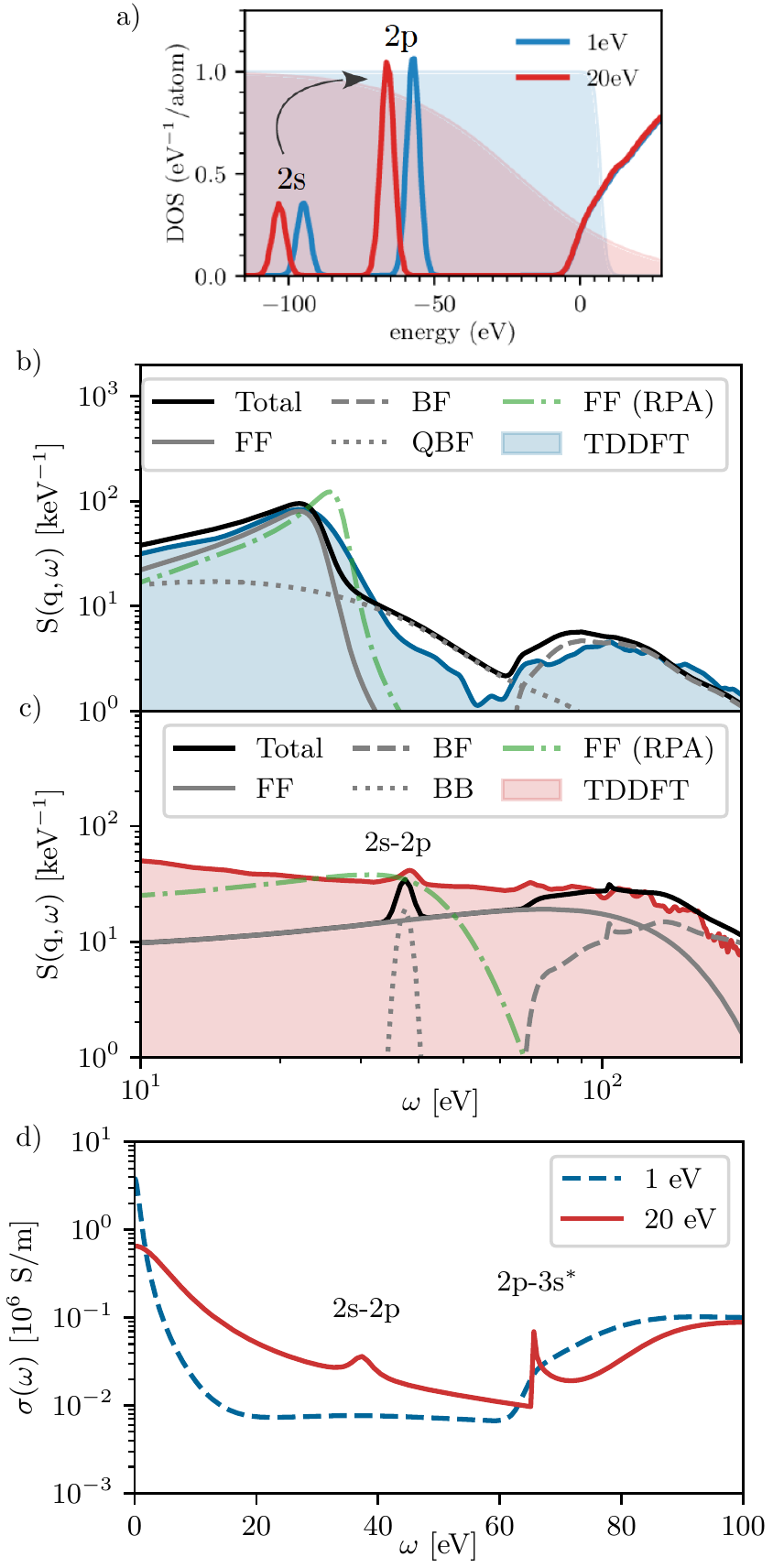}
    \caption{Bound-bound transitions in Al XRTS.
    a) DOS overlaid with the Fermi-Dirac distribution at \SI{2.70}{\gram\per\cm^3} and temperatures below (\SI{1}{\electronvolt}) and above (\SI{20}{\electronvolt}), where the 2p orbital becomes thermally depleted.
    b) DSF at \mbox{\SI{1}{\electronvolt}} and q=\SI{1.6}{\per\angstrom} computed using Chihara-decomposed AA (lines) and Chihara-free TDDFT (shaded).
    c) DSF at \SI{20}{\electronvolt} and q=\SI{4.4}{\per\angstrom} computed using AA and TDDFT, with a signature of a 2s-2p bound-bound transition now evident in both theories.
    In both DSFs we have included the FF component computed using the RPA to illustrate the significant impact that our choice of a Mermin dielectric function has.
    d) Optical conductivity computed using AA at 1 and \SI{20}{\electronvolt} in which bound-bound features are found at \SI{20}{\electronvolt}, and non-Drude behavior due to non-ideal continuum states is evident.
    }
    \label{fig:aluminum_overview}
\end{figure}

The first addition, the bound-bound term, is a fourth contribution to the typical three-term Chihara decomposition $S_{el}+S_{ff}+S_{bf}$ of the form
\begin{equation}
    S_{bb}=\sum \limits_{i,f} g_{i}  n(\varepsilon_i)\left(1-n(\varepsilon_f)\right) \mathcal{M}_{if}(q,\omega) \phi(\omega).
\end{equation}
Here the summation is taken over all possible initial ($i$) and final ($f$) bound states, with degeneracy $g_{i}$, Fermi-Dirac occupation $n(\varepsilon)$, and momentum-/energy-transfer-dependent scattering matrix element $\mathcal{M}_{if}$, and area-normalized line shape $\phi(\omega)$~\footnote{See Supplemental Material in final version for more details.}.
From the occupation factors alone it is evident that bound-bound transitions will contribute most strongly to scattering when the initial state is occupied and the final state is unoccupied.
In a degenerate system the former factor will be close to one for any sufficiently bound state, but for the latter to be non-zero (i.e., for a nominally bound state to be unoccupied) that state must be close to the chemical potential and/or the temperature must be comparable to its binding energy.


\begin{figure*}
    \centering
    \includegraphics[scale=0.8]{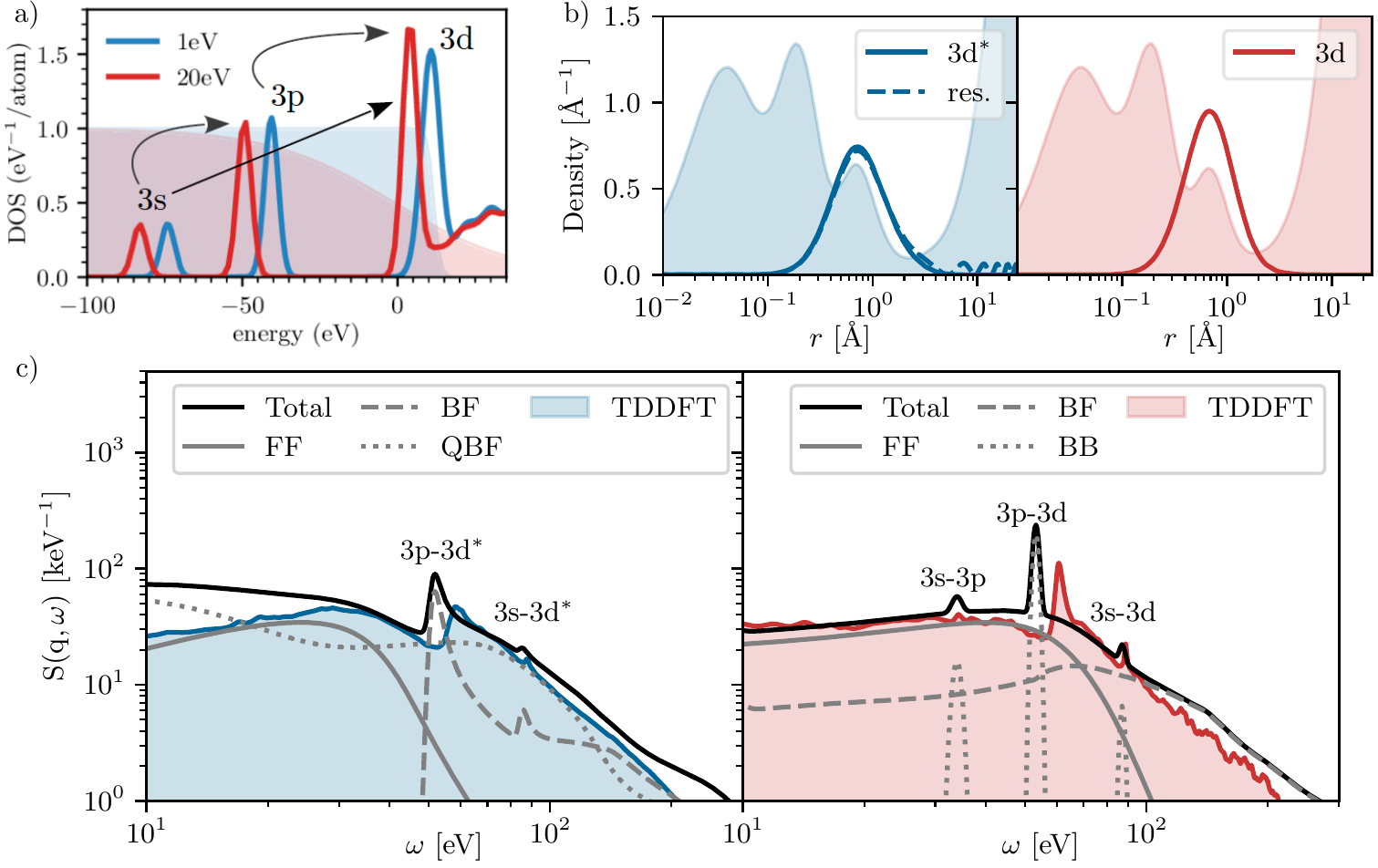}
    \caption{Bound-bound transitions in Fe XRTS.
    a) DOS overlaid with the Fermi-Dirac distribution at \SI{7.87}{\gram\per\cm^3} and temperatures below (\SI{1}{\electronvolt}) and above (\SI{20}{\electronvolt}) the point at which the 3p orbital will be thermally depleted and the 3d$^{*}$ quasibound state becomes bound 3d.
    b) The radial densities of the 3d$^{*}$ orbital and a d-like resonance (res.) in our AA model at \SI{1}{\electronvolt} (left) smoothly evolve into a 3d orbital at \SI{20}{\electronvolt} (right).
    The total densities of the Fe atom normalized by its nuclear charge are shown as shaded in the background.
    c) DSF at \SI{1}{\electronvolt} (left) and \SI{20}{\electronvolt} (right) for q=\SI{2.2}{\per\angstrom} computed using AA and TDDFT.
    At the lower temperature, the 3d$^{*}$ orbital is affiliated with two bound-quasibound features, while at the higher temperature, these become bound-bound features and 3p is thermally depleted to the point that a 3s-3p channel opens.
    }
    \label{fig:iron_overview}
\end{figure*}

Ensuring that states close to the chemical potential are treated continuously under modest changes in plasma temperature or density leads us to our second modification, a new definition of quasibound continuum states $n\ell^{*}$ as the ``scars'' of pressure-ionized bound states $n\ell$.
When a bound state moves into the continuum, its acquires a long-range oscillatory tail that resembles phase-shifted free-electron states, but retains significant bound-state character.
By averaging these orbitals over resonance features in the $\ell$-resolved density of states (DOS) we extract the bound-state character, generating quasibound orbitals whose statistical weights decrease as the associated resonances merge into the continuum~\footnote{See Supplemental Material in final version for more details.}.
These quasibound states are not included in $S_{bb}$, but are included in a second additional term, $S_{qbf}$, which captures transitions among distorted-wave states in the continuum and supplements the usual bound-free term given in Ref.~\cite{souza2014predictions}.
Together, the additional terms ensure continuity in the total DSF under pressure ionization of valence states.

Our final modification is to the free-free term. Here, we include only ideal free-electrons, with $\langle Z \rangle$ defined by the chemical potential $\mu$, and use a Mermin dielectric function \cite{mermin1970lindhard} with an improved treatment of the dynamic collision frequency, $\nu(\omega)$.
Based on T-matrix cross sections calculated using phase shifts from the self-consistent AA model, this model for $\nu(\omega)$ is more accurate than previous implementations that use a weak-scattering (Born) approximation.
This improved model tends to enhance shifts in the free-free term of the Chihara DSF that accompany the inclusion of the quasibound states in our reformulation of AA~\footnote{See Supplemental Material in final version for more details.}.

\textit{Results.---}

We first consider Al at ambient density and temperatures and momentum transfers indicated in Fig.~\ref{fig:aluminum_overview}.
It is worth noting that AA and DFT have good agreement in $\mu$ and the DOS when using the same exchange-correlation potential, and that while DOS and DSF data from the AA model are explicitly coupled to electronic orbitals, the DSF from TDDFT only implicitly depends on this DOS.
The 2s-2p bound-bound process indicated in the DOS is forbidden by Pauli exclusion at low temperatures, but becomes evident at \SI{20}{\electronvolt}, where the Fermi-Dirac occupation factors indicate significant vacancies in 2p.
Witte \emph{et al.} previously noted the appearance of a 2s-2p feature in the optical conductivity computed using the Kubo-Greenwood formalism at a temperature of \SI{12}{\electronvolt}~\cite{witte2018observations}.
Since the optical conductivity is related to the $q \rightarrow 0$ limit of the density response function, it is interesting to see whether this feature is predicted to manifest in inelastic scattering measurements that necessarily involve non-zero momentum transfer.
The DSFs under conditions in which the 2s-2p feature will not and will be evident are illustrated in panels b and c, respectively.
In both cases, the level of agreement between AA and TDDFT is similar to prior comparisons~\cite{baczewski2016x}.
Critically, both theories agree in their prediction of a bound-bound feature near \SI{35}{\electronvolt} (the energy separation of the bound states) at a temperature of \SI{20}{\electronvolt}.
Both TDDFT and AA models also show non-Drude features in the scattering signatures below the 60 eV L-edge at low temperatures, similar to those attributed to interband processes in low-q scattering experiments and calculated optical conductivities of Al at \SI{0.3}{\electronvolt}~\cite{witte2017warm}.
Panel d shows that optical conductivities from the AA model capture both the non-Drude character of Al and the 2s-2p transition, as well as a 2p-3s$^{*}$ bound-quasibound transition at \SI{20}{\electronvolt}.

We next consider Fe at ambient density and temperatures and momentum transfers indicated in Fig.~\ref{fig:iron_overview}.
Because of the presence of a narrow d-band resonance near the chemical potential, evident in the DsOS in panel a, we expect Fe to be an especially compelling test case for our AA model.
We expect to see bound-bound transitions involving the 3s, 3p, and 3d orbitals, with the d states taking on quasibound (3d$^{*}$) character at sufficiently low temperatures.
Panel b illustrates the transition from a quasibound to bound state under increasing temperature, as thermal excitations reduce the screening of the nuclear charge and the quasibound 3d$^{*}$ orbital and d-like resonance merge to become a bound 3d orbital.
At \SI{1}{\electronvolt} we expect to see transitions between 3s/3p and 3d$^{*}$, which are indeed evident in both AA and TDDFT (panel c).
Likewise at \SI{20}{\electronvolt} we expect to see bound-bound transitions along with a 3s-3p feature that is now enabled by thermal depletion of the 3p orbital.

\begin{figure}[h]
    \centering
    \includegraphics[scale=0.8]{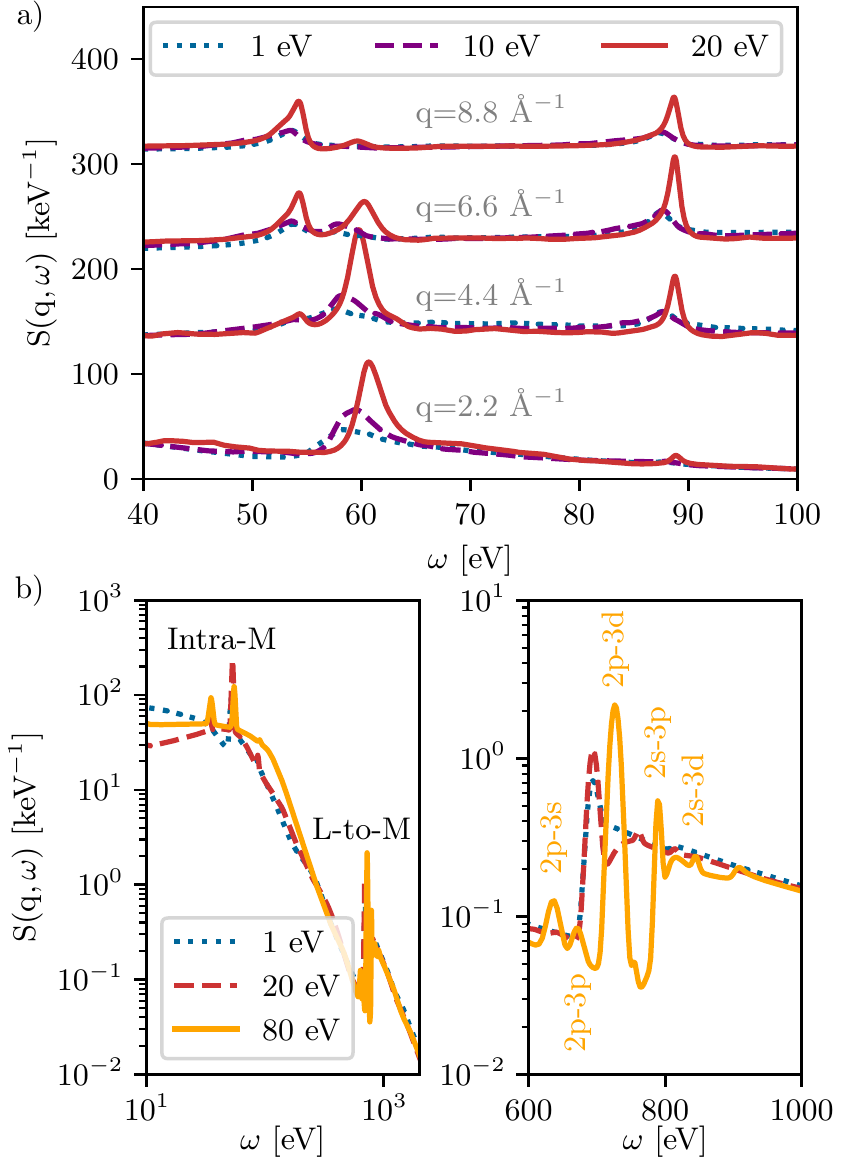}
    \caption{The impact of varying temperature on bound-bound transitions in iron.
    a) Varying both the temperature and momentum transfer using TDDFT, we find an unexpected collective character to the 3p-3d feature in the DSF.
    b) The AA DSF includes L-shell features that are challenging to capture using TDDFT and maintains computational efficiency even at higher temperatures (e.g., \SI{80}{\electronvolt}).
    This allows us to study the temperature dependence of bound-bound processes between the L and M shell, which may be used in thermometry.
    These data are for q=\SI{2.2}{\per\angstrom}.
    }
    \label{fig:thermal_effects}
\end{figure}

However, we note a discrepancy between AA and TDDFT in the energy transfers at which the 3p-3d$^{*}$ and 3p-3d features peak.
While the DFT and AA DsOS agree quite well, it is the AA model whose peak coincides with the energy difference between the 3p and relevant 3d levels, whereas in TDDFT this feature is higher in energy than indicated by a naive interpretation of the Kohn-Sham DOS.
This suggests that this particular feature has some collective character that lies beyond a simple picture in which the transition energies are strictly determined by energy differences between single-particle orbitals.
Indeed, this is precisely the sort of behavior that TDDFT excels at capturing.
We investigate this discordant feature and the broader impact of temperature on bound-bound features in Fig.~\ref{fig:thermal_effects}.

In Fig.~\ref{fig:thermal_effects}a, the DSF for Fe is computed at 3 different temperatures and 4 different momentum transfers.
We see that the energy of the 3s-3d feature changes by about \SI{2}{\electronvolt} in going from a temperature of $1$ to \SI{20}{\electronvolt} and that it does not disperse with increasing q, instead only varying in intensity.
However, while the energy of the 3p-3d feature above \SI{55}{\electronvolt} varies by approximately the same amount as a function of temperature, we note a slight but non-monotonic dispersion in q before it vanishes at high q.
More interestingly, as q increases the intensity of this feature is redistributed into a lower energy feature (below \SI{55}{\electronvolt}) at approximately the same energy as the AA 3p-3d$^{*}$ and 3p-3d features at \SI{54}{\electronvolt}.
This lends further credibility to the idea that this feature has some collective character that isn't captured in our AA model~\footnote{We also note that a feature at \SI{54}{\electronvolt} is evident in the Kubo-Greenwood optical conductivity, similar to the 2s-2p feature in Al.
However, Kubo-Greenwood also neglects collective effects, and thus the absence of a higher energy feature is consistent with our hypothesis.
This will also be further elaborated in the final version of this Letter.}.

This points to an open challenge pertaining to the treatment of states near the continuum in AA models.
Orbitals that are close to the chemical potential will often form narrow bands that are separated from bands with more conspicuous free-electron character (e.g., d- or f-electron metals).
While the quasibound treatment in this manuscript accounts for the bound character in a way that is continuous under ionization, it does not account for the collective excitations that might accompany the formation of bands comprised of these states.

While TDDFT captures effects that are inaccessible to the average atom model, the all-electron DFT-AA model can access high temperatures and inner-shell processes that are intractable for TDDFT.
Fig.~\ref{fig:thermal_effects}b shows the temperature dependence of the DSF around the L-edge of iron at temperatures ranging from 1 to \SI{80}{\electronvolt}.
Over this temperature range, the mean ionization increases from 1.8 to 8.2, decreasing the electronic screening of the central charge.
In response, the 2p and 3d states become more tightly bound, the energy of the L edge increases by \SI{90}{\electronvolt}, and the 2p-3d scattering signal changes smoothly from a sharp feature atop the L edge to a distinct bound-bound transition.
It is notable that this scattering feature appears at the same energies as RIXS features relative to the energy of incident photons.
While RIXS intensities dominate over scattering signals for on-resonance incident intensities characteristic of modern XFELS~\cite{humphries2020probing}, scattering signals persist for off-resonant and lower-intensity incident beams where RIXS signals are small.

\textit{Conclusion.---}

In this Letter we hypothesized the presence of bound-bound transitions that should be evident in XRTS experiments in laboratory-accessible warm dense conditions.
We find that both first-principles TDDFT and simplified AA models predict the emergence of these transitions, which convey a wealth of information about electronic structure and thermal effects that could be used in thermometry of warm dense matter. Inner-shell transitions into thermally depleted valance states may be particularly rich as diagnostics.

Although both AA and TDDFT models predicted the emergence of bound-bound features at temperatures near the Fermi energy in solids, these signatures of bound-bound scattering are inherently included in TDDFT, while the AA formalism required significant modifications.
In particular, AA required a new treatment of weakly bound orbitals that can move into the continuum with changes in temperature and density and attendant additional terms in the Chihara decomposition of the DSF.
These improvements to AA modeling will greatly benefit experiment as these methods are efficient enough to be employed in rigorous inference and can access inner-shell processes and temperatures that would be prohibitively expensive for TDDFT.

While TDDFT carries significant computational cost, it is absolutely critical as a benchmark for simpler models in warm dense conditions.
This is nicely exemplified by the discordance between AA and TDDFT for the 3p-3d feature in Fe scattering.
TDDFT suggests that some accommodation needs to be made for collective character in that feature, pointing to future work in AA method development.

\begin{acknowledgments}
We gratefully acknowledge useful conversations with
Attila Cangi, 
Kyle Cochrane, 
Laura Johnson, 
Walter Johnson, 
Hae Ja Lee, 
Charlie Starrett, 
Brian Wilson, 
and Bastian Witte. 
All authors contributed to the theory, analysis, and writing of this manuscript.

SBH and TH were partially supported by the US Department of Energy, Office of Science Early Career Research Program, Office of Fusion Energy Sciences under Grant No. FWP-14-017426.
ADB and AK were supported by the US Department of Energy Science Campaign 1.
All authors were partially supported by Sandia National Laboratories' Laboratory Directed Research and Development Program.

Sandia National Laboratories is a multi-mission laboratory managed and operated by National Technology and Engineering Solutions of Sandia, LLC, a wholly owned subsidiary of Honeywell International, Inc., for DOE's National Nuclear Security Administration under contract DE-NA0003525.
This paper describes objective technical results and analysis.
Any subjective views or opinions that might be expressed in the paper do not necessarily represent the views of the U.S. Department of Energy or the United States Government.
\end{acknowledgments}

\end{document}